\documentclass[12pt,epsf,epsfig,psfig]{article}
\usepackage{graphicx}
\usepackage{epsfig}
\def\e{\epsilon}

\def\be{\begin{equation}}
\def\ee{\end{equation}}

\def\lsim{\raise0.3ex\hbox{$<$\kern-0.75em\raise-1.1ex\hbox{$\sim$}}}
\def\gsim{\raise0.3ex\hbox{$>$\kern-0.75em\raise-1.1ex\hbox{$\sim$}}}

\def\NP{{ Nucl.\ Phys.\ }}
\def\PL{{ Phys.\ Lett.\ }}

\def\PRL{{ Phys.\ Rev.\ Lett.\ }}

\def\ZP{{ Z.\ Phys.\ }}

\begin{document}

\hfill BI-TP 2003/20

\vskip 1cm

\centerline{\Large \bf Conditions for Confinement and Freeze-Out}

\vskip 0.7cm

\centerline{V.\ Magas$^a$ and H.\ Satz$^{a,b}$}

\vskip 0.7cm

\centerline{a: Centro de Fis{\' \i}ca das Interac{\c c}{\~o}es
Fundamentais (CFIF)}

\centerline{Instituto Superior T\'ecnico, Av. Rovisco Pais, P-1049-001 Lisbon,
Portugal}

\medskip

\centerline{b: Fakult\"at fur Physik, Universit\"at Bielefeld}
\centerline{Postfach 100 131, D-33501 Bielefeld, Germany}

\vskip 1cm

\centerline{\bf Abstract:}

\bigskip

Matter implies the existence of a large-scale connected cluster of
a uniform nature. The appearance of such clusters as function of
hadron density is specified by percolation theory. We can
therefore formulate the freeze-out of interacting hadronic matter
in terms of the percolation of hadronic clusters. The resulting
freeze-out condition as function of temperature and baryo-chemical
potential interpolates between resonance gas behaviour at low
baryon density and repulsive nucleonic matter at low temperature,
and it agrees well with data.

\vskip 1cm

\begin{figure}[h]
\centerline{\hskip -0.4cm \psfig{file=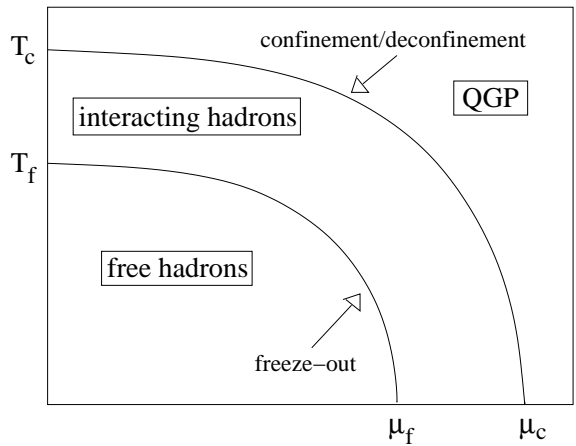,width=7.5cm}}
\caption{States of matter in QCD - general view}
\label{freeze}
\end{figure}

Consider a hot quark-gluon plasma in thermodynamic equilibrium,
specified in terms of temperature $T$ and baryochemical potential
$\mu$. Reducing the temperature of this medium at constant $\mu$
eventually brings it to the confinement transition at $T_c, \mu(T_c)$.
Below this temperature, the system consists of interacting hadrons.
Further cooling finally leads to freeze-out, at $T_f,\mu(T_f)$;
beyond this point, we have non-interacting hadrons. The resulting
two dividing lines in the $T-\mu$ plane are schematically shown in
Fig.\ \ref{freeze}.

\bigskip

While the line separating the deconfined QGP state from the confined
hadronic medium is calculable in lattice QCD at finite temperature and
density  (although so far with considerable problems at large $\mu$
and low $T$), the freeze-out curve is less well-defined theoretically
as well as experimentally. The aim of the present paper is to address
how freeze-out can be specified conceptually.

\medskip

In the two limiting cases of hot hadronic matter of vanishing overall
baryon density and cold nuclear matter at vanishing temperature, we can
make use of some specific aspects of the relevant dynamics. The confinement
transition leads to an interacting hadronic medium. Now it is known that
if the interactions between the constituents of such a medium are dominated
by resonance formation and decay, the system can be treated as an ideal
gas of all possible resonances \cite{B-U,Hagedorn}. In our context this
means that if and when the interacting hadronic medium is resonance
dominated, freeze-out effectively occurs at the point of confinement. In
particular, the relative abundances of the different species of hadrons
and hadronic resonances are in this case determined at the transition
from QGP to hadronic matter.

\medskip

For systems of vanishing or low baryon density, this appears quite
well supported. On a theoretical level, it is in fact claimed that
resonance formation dominates the interaction between hadrons, perhaps
most clearly in the dual resonance model \cite{DRM}. Experimentally,
an analysis  of the species abundances indicate that these are indeed
determined by a single freeze-out temperature, $T_f \simeq 175$ MeV,
obtained $e^+e^-$ annihilation, $p-p$ and $p-\bar p$ interactions as
well as in heavy ion collisions \cite{Becattini}. This temperature is
moreover completely in accord with that found for the confinement
transition in finite temperature lattice QCD \cite{lattice}.

\medskip

On the other extreme, for dense nuclear matter at low temperature, the
situation is quite different. The interaction between two nucleons does
not lead to resonance formation; instead, it is dominated by Fermi
statistics and baryon repulsion. In particular, at $T=0$, freeze-out
must occur at a density equal to that of normal nuclear matter,
$n_0 \simeq 0.17$ fm$^{-3}$. If we consider a system of non-interacting
nucleons at $T=0$, the corresponding value of the baryochemical potential
is obtained from
\be
n_0(T=0,\mu) = {2\over 3 \pi^2} (\mu^2 - m^2)^{3/2} = 0.17~{\rm fm}^3
\label{1}
\ee
and found to be $\mu_0 \simeq 0.979$ GeV. Note that for $T=0$
and this value of $\mu_f$, the system consists only of nucleons. The
threshold for the occurrence of baryon resonances is $\mu=1.23$ GeV,
corresponding to 4.2 fm$^{-3}$ or about 25 times normal nuclear density.
Hence interacting cold nuclear matter very likely contains only nucleons
over the entire range from freeze-out at $\mu_f = \mu_0$ GeV to the
deconfinement value $\mu_c$, where it becomes cold quark matter. The
freeze-out value of $\mu$ thus obtained is, however, in principle
somewhat model-dependent. We have here assumed the repulsion to be totally
given by Fermi statistics; there could well be additional repulsion
beyond this. Using hard core repulsion with a core volume $V_0$
\cite{CRSS}, Eq.\ \ref{1} is replaced by
\be
n_V(T=0,\mu) = {n_0(T=0,\mu) \over 1+ n_0(T=0,\mu)V_0} = 0.17~{\rm fm}^3.
\label{2}
\ee
The result is a freeze-out at slightly higher values of $\mu$. To illustrate,
for $V_0 = (4\pi /3)R^3$ and a nucleon radius of 0.8 fm, freeze-out
occurs at $\mu_V\simeq 0.992$ GeV. Such a hard core is presumably too large;
reducing $R$ will reduce the resulting $\mu_V$, so that the physical
value falls most likely between the values of $\mu_0$ and $\mu_V$.

\medskip

\begin{figure}[h]
\centerline{\hskip -0.7cm \psfig{file=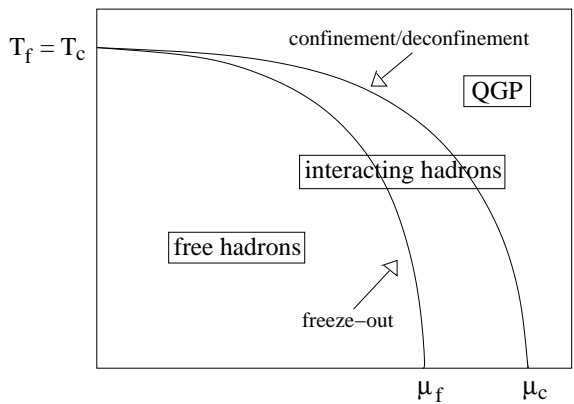,width=7.5cm}}
\caption{States of matter in QCD, including resonance gas effects}
\label{freeze4}
\end{figure}

Empirically, the determination of freeze-out parameters through the
relative abundance of hadron species so far appears to be the only
unambiguous approach to the problem. We therefore adopt this ``chemical''
freeze-out definition in the remainder of the paper. As an immediate
consequence, we have to revise Fig.\ \ref{freeze}. At $\mu=0$, $T_f=T_c$,
while at $T=0$, $\mu_f\simeq \mu_0 < \mu_c$, as shown in Fig.\
\ref{freeze4}. From what was said above, it is clear that for very
low temperatures, specifying freeze-out through species abundances
becomes problematic, since near $T=0$, the system contains only nucleons.
This has to be kept in mind when studying cool baryon-rich media.

\bigskip

Looking at Fig.\ \ref{freeze4}, we note that we have so far
identified only $T_f=T_c$ at $\mu=0$ and $\mu_f$ at
$T=0$. To determine the entire freeze-out curve in the $T-\mu$ plane,
we have to know how the contributions from non-resonant baryon
interactions modify the ideal resonance gas picture, in order to
specify how the freeze-out curve departs from the confinement curve.
In other words, we have to identify a freeze-out condition.

\medskip

A very successful phenomenological parametrization \cite{C-R} of the
available heavy ion data is obtained by requiring the average hadron
energy per average hadron number to be constant at freeze-out,
\be
{\langle E \rangle \over \langle N \rangle} \simeq 1~{\rm GeV}.
\label{3} \ee
While this parametrization accounts well for all
data from SIS to RHIC experiments, it is clear that it encounters
difficulties for sufficiently low temperature. At $T\!=\!0$, it
leads to freeze-out at $\mu \simeq 1.039$ GeV. With Fermi
repulsion only, this corresponds to $n_f \simeq 0.73~{\rm fm}^{-3}
\simeq 4.6~ n_0$, with the hard core repulsion used above it gives
$n_f \simeq 0.285~{\rm fm}^{-3} \simeq 1.7~n_0$. Both values would
exclude the existence of normal nuclear matter, for which $\langle
E \rangle /\langle N \rangle$ is below 1 GeV. Since it is also not
known what features of the underlying physics could lead to Eq.\
(\ref{8}) as freeze-out condition, it seems worthwhile to study
possible freeze-out mechanisms and attempt to find a consistent
description for the entire $T-\mu$ plane.

\medskip

Matter implies the existence of a large-scale interconnected
medium of uniform nature. When such a system breaks up into
fragments much smaller than the size of the volume in which it is
contained, it has undergone a change of state. One possible way to
define freeze-out is thus geometric: it occurs at the point at
which the size of the largest hadronic clusters falls below the
size of the given overall spatial volume. This point is determined
in percolation theory \cite{Isi} and for three space dimenensions
becomes
\be
n_f = {0.34 \over V_h},
\label{4}
\ee
where $n=N/V$ specifies the hadron density, with $N$ hadrons in the
overall volume $V$. The volume of an individual hadron is denoted by
$V_h=(4\pi/3)r_h^3$, and hadrons are allowed to overlap; thus $V_h$
introduces the short range nature ($\sim r_h$) of hadronic
forces \cite{perco}.

\medskip

When the hadron density has reached the percolation point $n=n_f$,
the largest clusters reach the size of the overall volume. This,
however, does not imply that they fill the volume. In fact, it is known
that at the percolation point, still $\exp\{-0.34\} \simeq 71$\% remain
empty space. Hence the vacuum, measured in terms of the hadronic scale
$r_h$, also forms percolating clusters. We can therefore ask for what
density vacuum percolation stops and only the strongly interacting medium
spans the entire space. This occurs for
\be
n_c = {1.24 \over V_h},
\label{5}
\ee
obtained in an analogous way as Eq.\ \ref{1}. For densities above
$n=n_c$, any large-scale vacuum has disappeared. Since the disappearence
of the physical vacuum is a basic feature of deconfinement, it seems
natural to relate this threshold to the confinement/deconfinement
transition.

\medskip

Hence, on a purely geometric basis in terms of connected clusters,
we have two thresholds: at $n=n_f$, there appear clusters of the
size of the overall spatial volume, and at $n=n_c$, the vacuum
disappears as a connected medium \cite{perco}. Let us see if and
how they be related to the known freeze-out points at $T=0$ and at
$\mu=0$.

\medskip

At $T=0$, we expect freeze-out to occur only when the nucleons no longer
form interconnected matter. If we use the nucleon radius $r_n \simeq 0.8$
fm for $r_h$, Eq.\ (\ref{4}) leads to $n_f \simeq 0.16$ fm$^{-3}$ as the
freeze-out density, and this is in good approximation the density of normal
nuclear matter. The correspondence between hadron percolation and freeze-out
thus works well at $T=0$. It defines nuclear matter as the most dilute
system of nucleons which still forms connected matter.

\medskip

Can we also relate the other extreme, freeze-out and deconfinement at
$\mu=0$, to percolation? Eq.\ (\ref{5}) determines the density of hadrons
for which the vacuum disappears as a connected medium. From the arguments
given above, we can consider the interacting hadron system at $\mu=0$ as
an ideal resonance gas. Hence it seems natural to use the vacuum percolation
condition
\be
n(T,\mu=0) = {1.24 \over (4\pi/3) r_h^3},
\label{6}
\ee
for the ideal resonance gas density $n(T,\mu=0)$ to determine
both freeze-out and deconfinement. To obtain $n(T,\mu)$ and
other necessary observables of such a system, we briefly recall
the essentials of the ideal resonance gas model.

\medskip

In thermodynamics, a closed system of fixed energy $E$, volume $V$,
and number $N_i$ of particles of type $i$ is described as microcanonical
ensemble. Bringing this system in contact with a heat bath leads to the
canonical ensemble, having an average energy $\langle E \rangle =
{\rm tr}(\rho H)$, while $V$ and $N_i$ are conserved exactly. If we
further also let the particle numbers $N_i$ fluctuate with only
their averages conserved, we obtain the grand canonical ensemble.
For this, we have
\be
Z_G = tr\, e^{-\beta(H - \mu_i N_i)}
= \prod_i \sum_{N_i} \lambda_i^{N_i} Z_{N_i},
\label{eq:zg}
\ee
where $\lambda_i = e^{\beta\mu_i}$ is the fugacity of particle species
$i$ and $Z_{N_i}$ the corresponding canonical partition function. The
chemical potentials $\mu_i$ assure particle number conservation in an
average sense. Changing the summation over discrete quantum states to
phase space integration, we obtain the grand canonical partition in the
form
\be
\ln Z_G =
\sum_i g_i {V  \over 2\pi^2} \int_0^{\infty}  dp\,  p^2 \ln\!
\left[ 1 + \eta_i \lambda_i e^{-\beta p_i^0}\right]^{\eta_i},
\ee
where $p_i^0=\sqrt{p^2+m_i^2}$, with $m_i$ for the mass of the particle
species $i$, $g_i$ counts the spin/isospin degeneracy, and $\eta_i$ is
the statistics factor, with $\eta_i = 1$ for fermions and $\eta_i = -1$
for bosons.

\medskip

In a relativistic gas, in which the particle numbers are generally not
conserved, the chemical potentials are associated to the conserved
baryon, charge and strangeness quantum numbers  $B$, $Q$ and $S$.
The abundance of strangeness is complicated by the larger strange
quark mass, which could prevent that strange quarks are produced
as abundantly as the lighter non-strange quarks. In order to allow
for a resulting reduced strangeness production, the parameter
$\gamma_s$ ($\gamma_s\le 1$) was introduced \cite{gammaS}.
For $\gamma_s=1$, no suppression occurs, while for $\gamma_s < 1$,
fewer strange quarks appear, with a corresponding suppression of
strange particle production. A factor $\gamma_s$ is associated to
each strange quark, so that for hadrons containing $f_s$ strange quarks,
there is an additional multiplier $\gamma_s^{f_s}$ in the fugacity.
We thus have
\be
\lambda_i = \gamma_s^{f_s}\lambda_B^{B_i}\lambda_Q^{Q_i}\lambda_S^{S_i}\,,
\label{eq190}
\ee
where $B_i$, $Q_i$ and $S_i$ are the quantum numbers for individual
particle species. Then the particle numbers are given by
\be
\langle N_i \rangle = g_i {V \over 2\pi^2} \int_0^{\infty} dp\,  p^2
\left[(\gamma_s^{f_s}\lambda_B^{B_i}\lambda_Q^{Q_i}\lambda_S^{S_i})^{-1}
e^{\beta p_i^0} + \eta_i\right]^{-1}\,,
\label{eq:numBSQ}
\ee
and the overall density of hadronic constituents by
\be
n(T,\mu) = \frac{\sum_i  \left \langle N_i \right \rangle}{V}\,.
\label{7}
\ee
The average net baryon number, i.e., the number of baryons minus that
of antibaryons, is the sum over the particle numbers weighted by the
baryon quantum number,
\be
\langle N_B \rangle = \sum_i B_i \langle N_i \rangle,
\label{8}
\ee
and the corresponding baryon density is
\be
n_B(T,\mu) = {\langle N_B \rangle \over V}.
\label{9}
\ee
From
\be
 \langle E \rangle  =
\sum_i g_i {V \over 2\pi^2} \int_0^{\infty} dp\,  p^2 p^0_i
\left[(\gamma_s^{f_s}\lambda_B^{B_i}\lambda_Q^{Q_i}\lambda_S^{S_i})^{-1}
e^{\beta p_i^0} + \eta_i\right]^{-1}
\label{10}
\ee
we obtain with
\be
\e=\frac{\langle E \rangle}{V}.
\label{11}
\ee
a similar relation for the energy density.

\medskip

For $\mu=0$, we now use Eq.\ (\ref{7}) in the percolation
condition (\ref{5}), including all observed resonances up to mass
2.5 GeV, with the nucleon radius $r_h=0.8$ fm for baryons; for
mesons, we use a value smaller by a factor $(2/3)^{3/4} \simeq
1.1$, as suggested by bag model arguments. This determines the
threshold temperature for confinement and (species abundance)
freeze-out at vanishing baryon-density, giving $T_c=T_f \simeq
167$ MeV for $\gamma_s=1$ and 175 MeV for $\gamma_s=0.5$.

\medskip

We have thus specified $T_f=T_c$ at $\mu=0$ through vacuum percolation
and $\mu_f$ at $T=0$ through nucleon percolation.

\medskip

To define a freeze-out curve in the whole $T-\mu$ plane, we have to
combine the resonance gas aspects at low baryon density with the
baryon repulsion nature at high baryon density. If we move from $\mu=0$
to finite $\mu$, the resulting medium is no longer an ideal gas, since
now baryon interactions are present, which are not accountable in terms
of resonances. To reach freeze-out, the system
has to expand and cool off enough to stop these non-resonant baryon
contributions. It is thus clear that for $\mu \not= 0$, freeze-out will
occur for $T_f < T_c$, below deconfinement. In particular, as is seen
from Fig.\ \ref{freeze4}, when we reach $\mu=\mu_f$, the interacting
hadronic medium formed at the confinement point still contains mesons; but
these have disappeared when the medium has cooled off enough to freeze
out all dynamical baryon repulsion, leaving cold nuclear matter
of standard density.

\medskip

Given the overall hadron density $n(T,\mu)$ and the net baryon
number density $n_B(T,\mu)$ through Eq'ns.\ (\ref{7}) and (\ref{9}),
respectively, we can specify a freeze-out temperature showing
the behaviour just outlined. We assume that
the sector of vanishing baryon density freezes out according to the
resonance gas approximation and by vacuum percolation, that of finite
baryon number according to baryon percolation, to obtain
\be
n(T,\mu) = {1.24 \over V_h}\left[ 1 - {n_B(T,\mu) \over n(T,\mu)}
\right] + {0.34 \over V_h}\left[ {n_B(T,\mu) \over n(T,\mu)}
\right] \label{12} 
\ee 
as the defining equation for the freeze-out
curve. It is clear that when $\mu=0$, we recover condition
(\ref{5}), while in the limit of a cold nucleon gas, with
$n/n_B=1$, we get back Eq.\ (\ref{4}). Moreover, it determines
freeze-out fully in terms of geometric clustering based on the
intrinsic hadronic scale. The model contains no adjustable
parameters, with the exception of a possible $\gamma_s < 1$. In
principle, $\gamma_s$ could depend on the initial collider energy 
or on $\mu$. However, variations of $\gamma_s$ have only little
effect on the resulting curve, as already noted for the limiting
case of $\mu=0$, and this effect decreases further for higher
$\mu$. Therefore we will perform our calculations for fixed
$\gamma_s=0.5$.

\medskip

\begin{figure}[h]
\centerline{\psfig{file=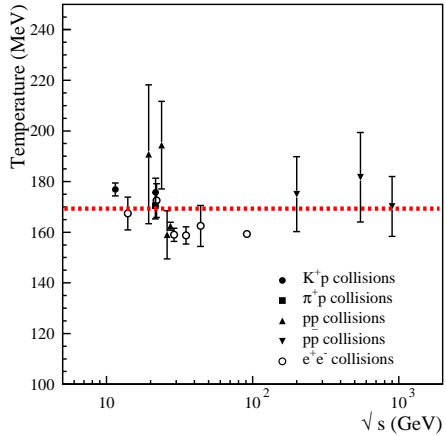,width=6.5cm}}
\caption{Freeze-out temperature in elementary hadron collisions, from
\cite{Becattini}.}
\label{EF1}
\end{figure}

We now turn to a comparison of our results to the relevant data.
Starting with data for freeze-out temperatures at $\mu=0$, we show
in Fig.\ \ref{EF1} a recent compilation of data from elementary
hadron collisions \cite{Becattini}; it is seen that the range $T
\simeq 170 \pm 5$ MeV obtained from the vacuum percolation
condition (\ref{4}) for $0.5 \leq \gamma_s \leq 1$ is in good
agreement with the data.

\begin{figure}[h]
\centerline{\hskip -0.7cm \psfig{file=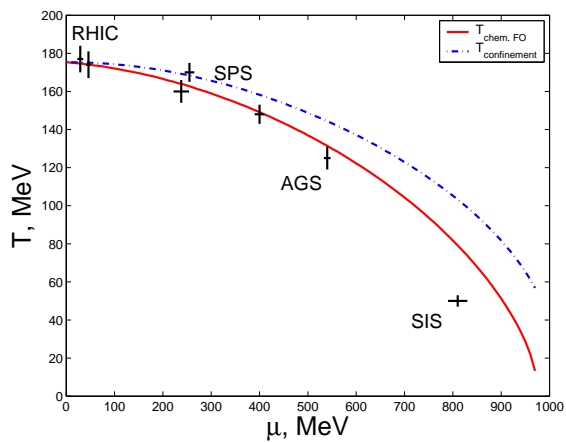,width=7.5cm}}
\caption{Freeze-out in heavy ion collisions, from \cite{KR}.}
\label{EF2}
\end{figure}

Coming to heavy ion collisions, we show in Fig.\ \ref{EF2} results
from the four different facilities. Also here the agreement is
seen to be very good, with clear deviations only for the lowest
energy data from SIS. While we have no explanation for this, it
should be noted that at such low energies, secondary hadron
production is a rather rare process, making a freeze-out
determination from species abundances more difficult. In
particular, the number of nucleons is now fixed by the mass of the
colliding nuclei, not by any thermal parameter. It is rather
surprising that the data fall below the expected curve. Our
assumption of an ideal nucleon gas at high baryon density could
well be an oversimplification; but we saw above that further
dynamical repulsion, beyond the effect due to the Fermi statistics
of the nucleons, in fact leads to a freeze-out at higher $\mu$.
One might thus wonder whether threshold effects at SIS energies
could lead to an effective reduction of the freeze-out
temperature.

\medskip

\begin{figure}[h]
\centerline{\psfig{file=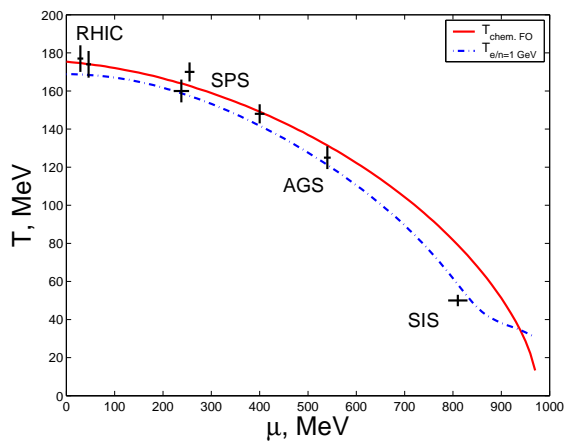,width=7.5cm}}
\caption{Freeze-out from percolation and from $\langle E \rangle /
\langle N \rangle = 1$  GeV.} \label{EF3}
\end{figure}

Finally, we compare our freeze-out condition with that
obtained from $<E>/<N> =1$ GeV \cite{C-R}. In Fig.\ \ref{EF3}, we show
both forms together with the data. While the two models never
differ by more than 10\%, the form of \cite{C-R}, in contrast to
ours, provides good agreement also for the SIS point. On the other
hand,  $<E>/<N> =1$ GeV leads to a rather sudden change of behaviour
for $\mu \geq 0.85$ GeV, which puts nuclear matter below the freeze-out
point in $\mu$.

\bigskip

\centerline{\bf Acknowledgments}

\medskip

We thank K.\ Redlich (Bielefeld and Wroclaw) for helpful
discussions and J.\ Manninen (Oulu) for valuable assistance with
programing.

\bigskip


\end{document}